\documentclass{pasa}%

\title[X-ray binary astrometry with VLBI]{Astrometric observations of X-ray binaries using very long baseline interferometry}
\author[James C.~A. Miller-Jones]{James C.~A. Miller-Jones\\
\affil{$^1$International Centre for Radio Astronomy Research - Curtin University, GPO Box U1987, Perth, WA 6845, Australia}}%
\jid{PASA}
\doi{10.1017/pas.\the\year.xxx}
\jyear{\the\year}

\usepackage{subfigure}
\usepackage{epsfig}
\usepackage[authoryear]{natbib}
\bibpunct{(}{)}{;}{a}{}{,}
\let\jnl@style=\rm
\def\ref@jnl#1{{\jnl@style#1}}

\def\aj{\ref@jnl{AJ}}                   
\def\actaa{\ref@jnl{Acta Astron.}}      
\def\araa{\ref@jnl{ARA\&A}}             
\def\apj{\ref@jnl{ApJ}}                 
\def\apjl{\ref@jnl{ApJ}}                
\def\apjs{\ref@jnl{ApJS}}               
\def\ao{\ref@jnl{Appl.~Opt.}}           
\def\apss{\ref@jnl{Ap\&SS}}             
\def\aap{\ref@jnl{A\&A}}                
\def\aapr{\ref@jnl{A\&A~Rev.}}          
\def\aaps{\ref@jnl{A\&AS}}              
\def\azh{\ref@jnl{AZh}}                 
\def\baas{\ref@jnl{BAAS}}               
\def\bac{\ref@jnl{Bull. astr. Inst. Czechosl.}}
\def\caa{\ref@jnl{Chinese Astron. Astrophys.}}
\def\cjaa{\ref@jnl{Chinese J. Astron. Astrophys.}}
\def\icarus{\ref@jnl{Icarus}}           
\def\jcap{\ref@jnl{J. Cosmology Astropart. Phys.}}
\def\jrasc{\ref@jnl{JRASC}}             
\def\memras{\ref@jnl{MmRAS}}            
\def\mnras{\ref@jnl{MNRAS}}             
\def\na{\ref@jnl{New A}}                
\def\nar{\ref@jnl{New A Rev.}}          
\def\pra{\ref@jnl{Phys.~Rev.~A}}        
\def\prb{\ref@jnl{Phys.~Rev.~B}}        
\def\prc{\ref@jnl{Phys.~Rev.~C}}        
\def\prd{\ref@jnl{Phys.~Rev.~D}}        
\def\pre{\ref@jnl{Phys.~Rev.~E}}        
\def\prl{\ref@jnl{Phys.~Rev.~Lett.}}    
\def\pasa{\ref@jnl{PASA}}               
\def\pasp{\ref@jnl{PASP}}               
\def\pasj{\ref@jnl{PASJ}}               
\def\rmxaa{\ref@jnl{Rev. Mexicana Astron. Astrofis.}}%
\def\qjras{\ref@jnl{QJRAS}}             
\def\skytel{\ref@jnl{S\&T}}             
\def\solphys{\ref@jnl{Sol.~Phys.}}      
\def\sovast{\ref@jnl{Soviet~Ast.}}      
\def\ssr{\ref@jnl{Space~Sci.~Rev.}}     
\def\zap{\ref@jnl{ZAp}}                 
\def\nat{\ref@jnl{Nature}}              
\def\iaucirc{\ref@jnl{IAU~Circ.}}       
\def\aplett{\ref@jnl{Astrophys.~Lett.}} 
\def\apspr{\ref@jnl{Astrophys.~Space~Phys.~Res.}}
\def\bain{\ref@jnl{Bull.~Astron.~Inst.~Netherlands}} 
\def\fcp{\ref@jnl{Fund.~Cosmic~Phys.}}  
\def\gca{\ref@jnl{Geochim.~Cosmochim.~Acta}}   
\def\grl{\ref@jnl{Geophys.~Res.~Lett.}} 
\def\jcp{\ref@jnl{J.~Chem.~Phys.}}      
\def\jgr{\ref@jnl{J.~Geophys.~Res.}}    
\def\jqsrt{\ref@jnl{J.~Quant.~Spec.~Radiat.~Transf.}}
\def\memsai{\ref@jnl{Mem.~Soc.~Astron.~Italiana}}
\def\nphysa{\ref@jnl{Nucl.~Phys.~A}}   
\def\physrep{\ref@jnl{Phys.~Rep.}}   
\def\physscr{\ref@jnl{Phys.~Scr}}   
\def\planss{\ref@jnl{Planet.~Space~Sci.}}   
\def\procspie{\ref@jnl{Proc.~SPIE}}   

\begin{document}%
\begin{abstract}
I review the astrophysical insights arising from high-precision astrometric observations of X-ray binary systems, focussing primarily (but not exclusively) on recent results with very long baseline interferometry.  Accurate, model-independent distances from geometric parallax measurements can help determine physical parameters of the host binary system and constrain black hole spins via broadband X-ray spectral modelling.  Long-term proper motion studies, combined with binary evolution calculations, can provide observational constraints on the formation mechanism of black holes.  Finally, the astrometric residuals from parallax and proper motion fits can provide information on orbital sizes and jet physics.  I end by discussing prospects for future progress in this field.
\end{abstract}

\maketitle%
\section{Introduction }
\label{sec:intro}

The past decade has ushered in an epoch of precision astrometry, with increases in sensitivity and enhanced processing techniques permitting very long baseline interferometers to make astrometric measurements accurate to a few tens of microarcseconds.  This enables the measurement of model-independent parallax distances for radio-emitting objects out to several kiloparsecs, and proper motions for radio sources anywhere in the Galaxy \citep[e.g.][]{Bru11,Loi11}, and, over a sufficiently long time baseline, out to Local Group objects \citep{Bru05,Bru07}.

As Galactic objects with radio-emitting jets, X-ray binaries provide a potential set of astrometric targets that can be used to study jet physics and the formation of compact objects, and for which geometric parallax distances can be invaluable in constraining fundamental system parameters such as peak luminosity (relative to the Eddington luminosity) and black hole spin.  However, the radio emission from X-ray binaries depends strongly on the X-ray spectral state \citep[see, e.g.][for a review]{FBG04}, and is not always suitable as an astrometric target.  The radio emission at any particular wavelength is brightest at the peak of sporadic (and unpredictable) outbursts, and typically arises from relativistically-moving jet knots that are no longer causally connected to the binary system \citep[e.g.][]{Mir94}.  The radio emission is then quenched by a factor of at least several hundred \citep{Rus11} during the soft, thermal-dominant X-ray state seen following the peak of the outburst.  Except for the few outbursts without an associated ejection event \citep[e.g.][]{Rus12,Par13}, this implies that astrometric observations can only be carried out in the hard and quiescent states, when the faint, flat-spectrum radio emission is believed to arise from a relatively steady, compact jet that is causally connected to the binary system\citep[e.g.][]{Fen01}.

In this article, I review existing astrometric observations of X-ray binaries, giving an overview of astrometric VLBI techniques (Section~\ref{sec:methods}), and then focussing on the physical insights that can be derived from both geometric distances (Section~\ref{sec:distances}) and proper motion measurements (Section~\ref{sec:pms}).  I examine the use of astrometric residuals to determine orbital parameters or constrain jet sizes (Section~\ref{sec:residuals}), and finish with a discussion of the prospects for future progress in this field (Section~\ref{sec:future}), including not only recent developments in VLBI and the potential contribution of the Square Kilometre Array (SKA), but also the recent launch of the space-based optical astrometric mission, GAIA \citep{Per01}.

\section{Astrometric techniques}
\label{sec:methods}

The theoretical astrometric precision of an interferometer is given by the instrumental resolution divided by twice the signal-to-noise ratio of the detection.  For maximum baselines of several thousand kilometres (as for the Very Long Baseline Array or the European VLBI Network) and observing frequencies of a few GHz, then the maximum resolution is on the order of a milliarcsecond.  Thus, with a signal-to-noise of 10--20, we can achieve astrometric accuracies of a few tens of microarcseconds.  However, in typical astrometric VLBI experiments, these measured positions are not absolute, but measured relative to a nearby (typically extragalactic) background source, in a technique known as phase referencing \citep[see][for a detailed overview of astrometric techniques]{Fom95}.  With sufficient signal-to-noise, the final astrometric precision becomes limited not by statistical uncertainty, but by systematics introduced in interpolating the phase solutions from the nearby phase reference calibrator to the target.  Such systematic errors scale linearly with angular separation between science target and calibrator source \citep{Pra06}, and typically limit the achievable astrometric accuracy to a few tens of microarcseconds.

A typical phase referencing experiment will cycle continuously between a bright, stationary extragalactic background source and the science target of interest.  This not only increases the possible integration time beyond the atmospheric coherence time (allowing observations of weak targets), but provides a relative position for the science target relative to that assumed for the calibrator source \citep{Wro00}.  Successful phase transfer depends on reliably connecting the phases between adjacent scans on the calibrator source (i.e.\ sufficiently short cycle times) and on the accuracy of the interpolation (i.e.\ a sufficiently small angular separation between calibrator and target source).  Specialised calibration techniques \citep[recently reviewed by][]{Rei13} can be employed to remove uncorrected tropospheric and clock errors from the correlated data using geodetic blocks \citep[occasional short observations of multiple bright calibrators located across the entire sky;][]{Mio04}, or to account for tropospheric phase gradients by observing multiple calibrators close to the target source \citep{Fom05,Fom05b}.

Following the transfer of the phases from the nearby calibrator source, the target position may be determined (prior to performing any self-calibration) by fitting a model (typically a point source or a Gaussian) in either the $uv$- or the image-plane.  When imaging, care must be taken with data weighting in the case of an inhomogeneous array.  For relatively faint (sub-mJy) sources such as X-ray binaries in the hard or quiescent state, the astrometric accuracy is limited by sensitivity, implying a preference for natural over uniform weighting despite the consequent loss in resolution.  However, natural weighting can lead to the measurements being dominated by any systematics affecting the most sensitive baseline(s), so unless required for a detection of the target source, a mild down-weighting of the most sensitive antennas can provide better astrometric accuracy, even at the expense of some signal-to-noise \citep[see the discussions in, e.g.][]{Del09,Mil13}. Except at the peak of the hard state, these sources tend to be unresolved (even with VLBI), and introducing a Gaussian taper into the weighting function is typically unnecessary.

Before fitting for the astrometric parameters, the magnitude of the systematic uncertainties must be assessed, and various approaches have been presented in the literature.  The simplest is to make occasional observations of an astrometric check source, calibrated in an identical fashion to the science target.  Scaling the scatter in its measured positions by its angular separation from the calibrator (relative to that for the target) gives a rough estimate of the magnitude of the systematic errors.  Alternatively, for sufficiently bright sources, intra-epoch systematics can be estimated from the positional scatter between different frequency sub-bands \citep{Del09}.  Alternatively, the full set of measured positions from each frequency sub-band and each epoch can form the sample for a Monte Carlo bootstrapping method of determining the systematics \citep{Cha09}.  Finally, the systematics in both right ascension and declination can be adjusted until the final reduced $\chi^2$ value of the astrometric fit reaches 1 \citep[e.g.][]{Del09,Rei11}.

\begin{figure}
\begin{center}
\includegraphics[width=\columnwidth]{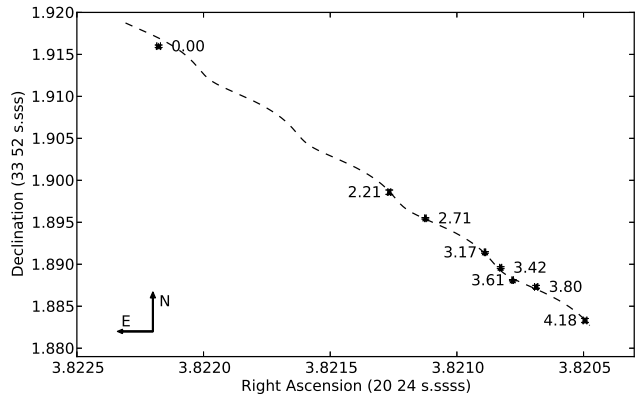}
\caption{Astrometric measurements of V404 Cygni over a period of over 4 years (the time of each epoch is marked on the trace, in years since the first observation).  The overall motion is to the southwest, with an annual parallax signature superposed.  Deconvolving these two signals allows a measurement of both the parallax and proper motion of the system. Adapted from fig.~1 of \citet{Mil09} by permission of the AAS.}\label{fig:v404}
\end{center}
\end{figure}

The five basic astrometric parameters are the source reference position and proper motion (in both right ascension and declination; $\alpha_0$, $\delta_0$, $\mu_{\alpha}\cos\delta$, $\mu_{\delta}$), and the source parallax ($\pi$).  Following \citet{Loi07}, the position of a source may be expressed in terms of these five parameters as
\begin{equation}
\begin{split}
\alpha(t)&=\alpha_0+(\mu_{\alpha}\cos\delta)t+\pi f_{\alpha}(t)\\
\delta(t)&=\delta_0+\mu_{\delta}t +\pi f_{\delta}(t),
\end{split}
\end{equation}
where $f_{\alpha}$ and $f_{\delta}$ are the projections of the parallax ellipse onto the right ascension and declination axes \citep{Sei92}.  This set of coupled equations can be solved using a singular value decomposition algorithm \citep[see][for details]{Loi07}.  The solution provides the reference position of the source at a given epoch, its proper motion and its parallax, from which the motion of the source on the sky can be determined, as shown in Figure~\ref{fig:v404}.

\section{The X-ray binary distance scale}
\label{sec:distances}

Distance is a fundamental quantity in astrophysics.  Accurate distances are required to convert observational quantities (such as measured fluxes and proper motions) into the corresponding physical quantities (luminosities and speeds, respectively).

Relying on simple geometry alone, trigonometric parallax is the only model-independent method of distance determination, and as such, is the gold standard against which we can calibrate all other methods.  However, X-ray binaries are not typically amenable to parallax measurements.  Other than a handful of Be/X-ray binaries \citep{Che98}, all known systems are located at $\geq 1$\,kpc; the closest known transient neutron star and black hole X-ray binaries are Cen X-4 \citep[$\leq1.2\pm0.3$\,kpc;][]{Che89} and A0620-00 \citep[$1.06\pm0.12$\,kpc;][]{Can10}, respectively.  Thus, in the majority of cases, the amplitude of the parallax signal would be less than 1\,mas.  While this level of accuracy can be achieved with Very Long Baseline Interferometry (VLBI), it requires the source to emit bright, compact radio emission.

The first reported VLBI parallaxes for X-ray binaries were for the two high-mass systems Cygnus X-1 and LSI +61$^{\circ}$303 \citep{Les99}, as part of a program to tie the Hipparcos optical reference frame to the International Celestial Reference Frame (ICRF).  While the post-fit residuals were too large to determine the distance to LSI +61$^{\circ}$303, the parallax of Cygnus X-1 was detected at the $2.4\sigma$ level, giving a distance of $1.4^{+0.9}_{-0.4}$\,kpc.

\subsection{Reaching the Eddington luminosity}

The first truly precise X-ray binary parallax measurement was made for the Z-source neutron star system Sco X-1 \citep{Bra99}.  Taking advantage of a nearby (70$^{\prime\prime}$) calibrator to minimise the astrometric systematics, the fitted parallax of $360\pm40$\,$\mu$as was at the time both the smallest and most precise measurement ever made.  The derived distance of $2.8\pm0.3$\,kpc proved that Sco X-1 did indeed reach its Eddington luminosity in a particular X-ray spectral state (the vertex of the normal and flaring branches in an X-ray colour-colour diagram). This work was also notable in demonstrating that despite amplitude and structural variations on timescales as short as 10 minutes, the radio core (assumed to correspond to the binary system itself) could be unambiguously identified in most epochs, thus permitting the required high-precision astrometric measurements.

Distances to neutron star X-ray binaries are often estimated by assuming that certain Type I X-ray bursts \citep[arising from unstable thermonuclear burning of accreted material; see][for a review]{Gall08}, known as photospheric radius expansion (PRE) bursts, reach the local Eddington luminosity, so can act as standard candles \citep[to within $\sim15$\%;][]{Kuu03}.  However, the expected peak luminosity varies according to the hydrogen mass fraction of the accreted material.  The Eddington luminosity for pure He burning is 1.7 times higher than for material of solar composition, corresponding to a 30\% change in the estimated distance.  Hence, with a sufficiently accurate distance measurement, it could be possible to determine the composition of the accreting material.  However, other sources of systematic error could blur this signal, arising from uncertainties in the neutron star mass \citep[which can be at least as high as $2M_{\odot}$;][]{Dem10}, the maximum radius reached by the expanding photosphere during the burst (affecting the gravitational redshift and hence the luminosity), and the 5--10\% variation in burst luminosities observed within a given source \citep{Gall08}.

Although Sco X-1 has never shown Type I X-ray bursts, there are a handful of other neutron star X-ray binaries showing both Type I bursts and detectable radio emission, at least at certain phases of their outburst cycles.  Of these, the most nearby sources (with the largest parallax signatures) are the ultracompact system 4U\,0614+091 \citep{Kuu10,Mig10}, and the atoll sources Aql X-1 \citep{Koy81,Mil10} and 4U\,1728-34 \citep{Hof76,Mig03}.  Parallax distances to these sources would extend the sample of systems used to calibrate the relationship between PRE burst luminosities and the Eddington luminosity \citep[previously restricted to 12 globular cluster sources;][]{Kuu03}.

Accurate estimates of the peak outburst luminosity are also important for Galactic black hole X-ray transients, since they can in principle shed light on the nature of ultraluminous X-ray sources \citep[ULXs; see][for a review]{Fen11}, particularly at the low-luminosity end of the ULX luminosity function, around $10^{39}$\,erg\,s$^{-1}$.  Although \citet{Gri02} found only 2 transient black hole X-ray binaries in the Galaxy to have exceeded $10^{39}$\,erg\,s$^{-1}$ (the Eddington luminosity for a $10M_{\odot}$ black hole), \citet{Jon04} demonstrated that with more accurate distance estimates, at least five, and possibly up to seven of the fifteen transient Galactic systems could have exceeded this limit, and would thus have been classified as ULXs had they been observed in an external galaxy.  Although the majority of ULXs are persistent rather than transient \citep{Fen11}, the lack of a break in the ULX luminosity function below $10^{40}$\,erg\,s$^{-1}$ suggests that stellar-mass black holes can indeed exceed the Eddington limit by at least a small factor, a hypothesis borne out by the recent detection of Eddington-rate behaviour in an outburst of a microquasar in our neighbouring galaxy, M31 \citep{Mid13}.

To date, only one transient Galactic black hole, V404 Cygni, has an accurate parallax distance measurement \citep{Mil09}.  This revised the source distance downwards by a factor of 1.7, implying that its 1989 outburst only reached a luminosity of $\sim0.5 L_{\rm Edd}$ \citep[see][for a detailed description of this outburst]{Tan95}.  While additional parallax measurements for transient sources would be valuable, the short (months-long) durations of their outbursts and the low quiescent luminosities of many systems \citep{Gal08,Mil11} preclude radio detections for all but the closest systems \citep[e.g.][]{Gal06}.  The best candidates for future parallax measurements would therefore be recurrent transients such as H1743-322 or GX339-4 (see Section~\ref{sec:future}).

\subsection{Evidence for event horizons}

Accurate luminosities are not only important for X-ray binaries in outburst, but also in quiescence.  Black hole X-ray binaries have been found to have systematically lower bolometric luminosities than neutron star systems with similar orbital periods \citep{Nar97,Men99,Gar01}, which was attributed to the existence of radiatively inefficient accretion flows (RIAFs) and event horizons in the black hole systems.  Although this interpretation has been challenged \citep{Cam00,Abr02}, and alternative explanations are possible (e.g.\ coronal emission; \citealt{Bil00}, energy being channeled into jets; \citealt*{Fen03}), the original claim also relies on accurate estimates of the source luminosities, and hence distances.

\subsection{Determining system parameters}

Accurate distances to black hole X-ray binaries can also be invaluable in constraining the physical parameters of the binary system.  In quiescent systems, an accurate distance can be used to determine the luminosity of the hotspot where the accretion stream impacts the disc, allowing the mass transfer rate from the secondary star to be determined \citep[e.g.][]{Fro11}, providing important observational constraints for binary evolution models.

Other key system parameters that can benefit from accurate distance determinations are the component masses and orbital inclination angle.  Using a larger bandwidth and an observational strategy designed to minimise systematic uncertainties (such as the use of geodetic blocks), \citet{Rei11} revisited the parallax of Cygnus X-1, finding a distance of $1.86^{+0.12}_{-0.11}$\,kpc; consistent with, but significantly more precise than the previous measurement of \citet{Les99}.  The source distance had been a major uncertainty in determining the system parameters of this persistent X-ray binary \citep[e.g.][]{Pac74}, which contains the first black hole to be discovered, and has since become one of the most well-studied black hole systems, providing important insights into accretion physics.

With the new parallax distance, accurate to 6\%, \citet{Oro11} were able to determine the donor star radius from its K-band magnitude, thereby strongly constraining the dynamical model for the system.  Adding in other constraints (radial velocity curves and optical photometry), they determined the black hole and donor masses, the inclination angle of the orbital plane, and measured a non-zero eccentricity for the orbit.

The uncertain distance for GRS\,1915+105 also provides the bulk of the uncertainty in determining its system parameters \citep{McC06}, which have recently been revised by \citet{Stee13}.  The latter authors are already undertaking an astrometric program to determine a parallax distance to the source, the results of which should finally pin down the nature of this enigmatic system, which has almost certainly been accreting close to the Eddington rate for over two decades, and has provided an ideal laboratory for studying disc-jet coupling \citep[see][for a review]{Fen04}. 

\subsection{Constraining black hole spin}

With accurate values of distance, inclination angle and black hole mass, it is possible to fit high-quality, disc-dominated X-ray spectra of black hole X-ray binaries with fully relativistic models for the accretion disc to measure the black hole spin \citep[see][for a review]{McC13}.  Since the black hole spin sets the radius of the innermost stable circular orbit (ISCO), then assuming that the accretion disc is sharply truncated at the ISCO, the spin can be determined from the fitted inner disc radius.  This method has so far been used to determine the spins of ten stellar-mass black holes.  However, it relies on accurate pre-existing constraints on the source distance, inclination angle and black hole mass, and uncertainties in these parameters are the major source of uncertainty in the derived spins.

Using the accurate values of distance \citep{Rei11}, inclination and mass \citep{Oro11} measured for Cygnus X-1, \citet{Gou11} were able to measure an extremely high value for the dimensionless spin parameter of $a_{\rm *}>0.95$, in good agreement with a recent measurement derived from an analysis of the relativistically-broadened Fe K$\alpha$ line profile \citep{Dur11}.  Such a high spin is believed to be shared only by GRS\,1915+105 \citep{McC06,Blu09} among the black hole X-ray binaries, and, if it can be tapped by the Blandford-Znajek mechanism \citep{Bla77}, implies the possibility of extremely powerful jets.

It has recently been claimed that ballistic jets from transient black hole X-ray binaries that reach a significant fraction of their Eddington limit are indeed powered by black hole spin \citep{Nar12,Ste13}.  This claim relies on an apparent correlation between the measured spins of selected transient black hole X-ray binaries and a proxy for their jet powers (the maximum unbeamed 5\,GHz radio luminosity during outburst, scaled by the black hole mass).  However, this remains controversial \citep{Fen10,Rus13}, owing to the difficulty in identifying an accurate proxy for the jet power, the inherent uncertainties on the measured black hole spins, and the small number of sources deemed to be suitable for inclusion in the sample.  More accurate distance measurements from VLBI parallaxes would help to reduce the uncertainties in the measured spins and jet powers, thereby helping to resolve this important debate.

\subsection{The neutron star equation of state}

As discussed by \citet{Tom09}, accurate distances to neutron star systems can also help constrain the neutron star equation of state \citep{Lat07}.  Different equations of state produce different mass-radius relationships, and although recent years have seen some extremely accurate neutron star mass measurements \citep[e.g.][]{Dem10}, radius determinations are often dependent on the source distance.  The quiescent X-ray emission from neutron star X-ray binaries is dominated by the blackbody emission from the neutron star surface.  A model-independent geometric distance measurement would allow the luminosity to be determined more accurately (to the accuracy of the X-ray flux scale, typically 10--20\%), allowing the area (and hence the radius) of the emitter to be determined via the Stefan-Boltzmann law.  A precise determination of both mass and radius for just a single neutron star would be invaluable in ruling out many of the proposed equations of state.

\section{Compact object formation and natal kicks}
\label{sec:pms}

Even for objects whose distances are too great, or for which systematic astrometric uncertainties are too large to measure a parallax distance, it is possible to measure a proper motion, since the signal is cumulative with time.  If the source distance and the systemic radial velocity can also be determined (the latter typically from optical or near-infrared spectroscopy), then all six position and velocity components are known.  By integrating backwards in time in the Galactic potential, it is possible to trace the orbit of the system through the Galaxy (e.g.\ Fig.~\ref{fig:potential}), and hence derive constraints on the formation of the compact object.

\subsection{Natal kicks}

\begin{table*}
\caption{Measured astrometric parameters of X-ray binaries.  Systems have been divided into confirmed black holes (top) and neutron stars (middle), and systems whose compact object is still unknown (bottom).} 
\begin{center}
\begin{tabular*}{\textwidth}{@{}c\x c\x c\x c\x c\x c\x c\x c@{}}
\hline \hline
Source & $l$ & $b$ & $d$ & $\mu_{\alpha}\cos\delta$ & $\mu_{\delta}$ & $\gamma$ & References\\
& ($^{\circ}$) & ($^{\circ}$) & (kpc) & (mas\,yr$^{-1}$) & (mas\,yr$^{-1}$) & (km\,s$^{-1}$) &\\
\hline
XTE J1118+480 & 157.66 & 62.32 & $1.72\pm0.10$ & $-16.8\pm1.6$ & $-7.4\pm1.6$ & $+2.7\pm1.1$ & [1,2,3] \\
GRO J1655-40 & 344.98 & 2.46 & $3.2\pm0.2^{a}$ & $-3.3\pm0.5$ & $-4.0\pm0.4$ & $-141.9\pm1.3$ & [4,5,6] \\
GRS 1915+105 & 45.37 & -0.22 & $11\pm1^{b}$ & $-2.86\pm0.07$ & $-6.20\pm0.09$ & $+11\pm4.5$ & [7,8] \\
Cyg X-1 & 71.33 & 3.07 & $1.86\pm0.12$ & $-3.78\pm0.06$ & $-6.40\pm0.12$ & $-5.1\pm0.5$ & [9,10]\\
V404 Cyg & 73.12 & -2.09 & $2.39\pm0.14$ & $-5.04\pm0.22$ & $-7.64\pm0.03$ & $-0.4\pm2.2$ & [11,12]\\
\hline
LSI +61$^{\circ}$303 & 135.68 & 1.09 & $2.0\pm0.2$ & $-0.30\pm0.07$ & $-0.26\pm0.05$ & $-40.2\pm1.9$ & [13,14,15]\\
Cen X-4 & 332.24 & 23.88 & $1.4\pm0.3$ & $-11\pm10$ & $-56\pm10$ & $189.6\pm0.2$ & [16,17]\\
Sco X-1 & 359.09 & 23.78 & $2.8\pm0.3$ & $-6.88\pm0.07$ & $-12.02\pm0.16$ & $-113.8\pm0.6$ & [18,19]\\
LS 5039 & 16.88 & -1.29 & $2.9\pm0.3$ & $7.10\pm0.13$ & $-8.75\pm0.16$ & $17.2\pm0.5$ & [20,21] \\
Aql X-1 & 34.67 & -4.68 & $5.0\pm0.9^{c}$ & $-2.64\pm0.14^{d}$ & $-3.53\pm1.40^{d}$ & $30\pm10$ & [22,23,24,25]\\
Cyg X-2 & 87.33 & -11.32 & $11\pm2^{c}$ & $-3.00\pm0.68$ & $-0.64\pm0.68$ & $-209.6\pm0.8$ & [24,26,27]\\
\hline
SS 433 & 39.69 & -2.24 & $5.5\pm0.2$ & -3.5 & -4.6 & $65\pm3$ & [28,29,30]\\
Cyg X-3 & 79.85 & 0.70 & $7.2^{+0.2}_{-0.5}$ & $-2.73\pm0.06$ & $-3.70\pm0.06$ & $<200$ & [31,32]\\
\hline \hline
\end{tabular*}\label{tab:pms}
\end{center}
\tabnote{$^a$This distance is derived from the proper motions of the relativistic jets, assuming an inclination angle for the system of $85\pm2^{\circ}$. Other authors have suggested a closer distance \citep{Mir02,Foe09}.}
\tabnote{$^b$Although a distance of $11\pm1$\,kpc is favoured should the systemic velocity track Galactic rotation, and also from the proper motions of relativistic jets, \citet{Kai04} have suggested distances as low as 6\,kpc.}
\tabnote{$^c$The quoted distance is for the accretion of material of solar metallicity onto the donor star; accretion of helium-rich material would give a distance higher by a factor 1.3 \citep{Gall08}.}
\tabnote{$^d$Proper motions have been deduced from the VLBI positions reported by \citet{Mil10} and \citet{Tud13}.}
\tabnote{References: [1] \citet{Mir01}; [2] \citet{Gel06}; [3] \citet{Gon08}; [4] \citet{Mir02}; [5] \citet{Hje95}; [6] \citet{Sha99}; [7] \citet{Dha07}; [8] \citet{Stee13}; [9] \citet{Rei11}; [10] \citet{Gie08}; [11] \citet{Mil09}; [12] \citet{Cas94}; [13] \citet{Dha06}; [14] \citet{Fra91}; [15] \citet{Cas05}; [16] \citet{Gon05}; [17] \citet{Cas07}; [18] \citet{Bra99}; [19] \citet{Ste02}; [20] \citet{Mol12a}; [21] \citet{Cas05a}; [22] \citet{Mil10}; [23] \citet{Tud13}; [24] \citet{Gall08}; [25] \citet{Cor07}; [26] \citet{Spe13}; [27] \citet{Ele09}; [28] \citet{Loc07}; [29] \citet{Blu04}; [30] \citet{Hil04}; [31] \citet{Mil09b}; [32] \citet{Lin09}.}
\end{table*}

The high space velocities of radio pulsars provide good evidence for strong natal kicks during the formation of neutron stars \citep{Lyn94}.  These kicks, which can give rise to velocities in excess of 1000\,km\,s$^{-1}$ \citep{Hob05}, cannot be explained purely by the supernova recoil kick \citep{Bla61}.  The recoil is set by the ejected mass, and since ejection of more than half the total mass causes a binary system to become unbound, this sets an upper limit to the maximum recoil velocity \citep{Nel99}.  Alternative possibilities for generating high natal kick velocities typically involve hydrodynamical mechanisms, asymmetric neutrino emission induced by strong magnetic fields, or electromagnetic kicks from an off-centre rotating dipole, and have been reviewed in detail by \citet{Lai01}.

A second population of neutron stars is believed to form with significantly lower kicks \citep{Pfa02a,Pfa02b}, potentially due a smaller iron core in the progenitor star, or to formation in an electron-capture supernova \citep{Pod04}.  The ensuing prompt or fast explosion does not allow time for convectively-driven instabilities to grow in the neutrino-heated layer behind the supernova shock \citep{Sch04}, leading to smaller kick velocities.  The recent discovery of two distinct subpopulations of Be/X-ray binaries provides further observational support for a dichotomy between these two types of supernova \citep{Kni11}.

Black holes are believed to form in two different ways \citep{Fry01}.  For a sufficiently massive progenitor, they may form by direct collapse.  Alternatively, if a supernova explosion is not sufficiently energetic to unbind the stellar envelope, fallback of ejected material onto the proto-neutron star formed in the explosion can create a black hole.  In the latter case, many of the non-recoil kick mechanisms that have been proposed for neutron stars (with the exception of the electromagnetic kicks) could also apply to black holes.

The similarity in the distributions of black hole and neutron star X-ray binary systems with Galactic latitude has been used to argue for equivalent natal kicks during black hole formation \citep{Jon04}.  Indeed, detailed population synthesis calculations have suggested (albeit discounting observational selection effects) that such kicks are necessary, with the magnitudes of black hole kick velocities (rather than their momenta) being similar to those of neutron stars \citep{Rep12}.  This latter point could be used to discriminate between proposed kick mechanisms; while neutrino-driven kicks should give rise to the same momenta in black holes and neutron stars, hydrodynamical kicks from asymmetries in the supernova ejecta can accelerate a nascent black hole to similarly high velocities as observed in neutron stars \citep{Jan13}.

Thus, VLBI measurements of the proper motions of black hole X-ray binaries can be used to probe the black hole formation mechanism, determining whether or not a natal kick is required for a given system, and, eventually, determining the distribution of black hole kick velocities.  A bimodal distribution would be good evidence for some black holes to form without a natal supernova, with the most massive black holes (not having lost material in the explosion) likely to have the lowest velocities relative to their local standard of rest (LSR).

\subsection{Observational constraints}

\begin{table*}
\caption{Inferred Galactic space velocities of X-ray binaries.} 
\begin{center}
\begin{tabular*}{\textwidth}{@{}c\x c\x c\x c\x c\x c\x c@{}}
\hline \hline
Source & $U$ & $V$ & $W$ & $UC$ & $VC$ & $v_{\rm pec}$\\
& (km\,s$^{-1}$) & (km\,s$^{-1}$) & (km\,s$^{-1}$) & (km\,s$^{-1}$) & (km\,s$^{-1}$) & (km\,s$^{-1}$)\\
\hline
XTE J1118+480 & $-94\pm13$ & $-90\pm14$ & $-20\pm6$ & $16$ & $-1$ & $144\pm14$\\
GRO J1655-40 & $-147\pm3$ & $-27\pm8$ & $2\pm7$ & $-39$ & $-3$ & $110\pm3$\\
GRS 1915+105 & $272\pm23$ & $-230\pm23$ & $-11\pm4$ & $238$ & $-228$ & $36\pm22$\\
Cyg X-1 & $72\pm4$ & $-14\pm1$ & $6\pm1$ & $55$ & $-6$ & $19\pm4$\\
V404 Cyg & $110\pm6$ & $-18\pm3$ & $4\pm1$ & $71$ & $-11$ & $40\pm6$\\
\hline
LSI +61$^{\circ}$303 & $41\pm1$ & $-15\pm1$ & $3\pm1$ & $35$ & $-3$ & $14\pm1$\\
Cen X-4 & $148\pm39$ & $-346\pm84$ & $-174\pm82$ & $-23$ & $-1$ & $422\pm78$\\
Sco X-1 & $-83\pm1$ & $-167\pm19$ & $-68\pm3$ & $-2$ & $0$ & $198\pm17$\\
LS 5039 & $42\pm2$ & $-42\pm6$ & $-136\pm15$ & $38$ & $-3$ & $141\pm14$\\
Aql X-1 & $82\pm17$ & $-36\pm23$ & $18\pm12$ & $99$ & $-22$ & $29\pm17$\\
Cyg X-2 & $142\pm44$ & $-184\pm7$ & $125\pm37$ & 196 & -103 & $159\pm33$\\
\hline
SS433 & $157\pm4$ & $-60\pm5$ & $30\pm3$ & 161 & -63 & $31\pm3$\\
\hline \hline
\end{tabular*}\label{tab:vpec}
\end{center}
\tabnote{$U$, $V$ and $W$ are defined as positive towards $l=0^{\circ}$, $l=90^{\circ}$, and $b=90^{\circ}$, respectively. $UC$ and $VC$ are the velocities expected from circular rotation at 238\,km\,s$^{-1}$ \citep{Hon12}.  The peculiar velocity $v_{\rm pec}$ is defined as $[(U-UC)^2+(V-VC)^2+W^2]^{1/2}$.}
\end{table*}

\subsubsection{Black holes}

The first evidence for a strong natal kick in a black hole system was found from optical spectroscopy of GRO J1655-40.  \citet{Bra95} considered possible explanations for the large measured systemic radial velocity of $150\pm19$\,km\,s$^{-1}$ \citep{Bai95}, including rocket acceleration by jets, a triple system, discrete scattering events, or perturbations due to interactions with density waves in the Galactic potential.  These were all deemed to be unlikely, leaving a natal kick in a supernova explosion as the most plausible explanation \citep[a scenario that is also supported by the observed misalignment between the disc plane and the jet axis;][]{Mac02}.  With the subsequent measurement of a proper motion for the system by the Hubble Space Telescope \citep{Mir02}, more detailed modelling was able to reconstruct the full evolutionary history of the binary system since the black hole was formed \citep{Wil05}.  Although formation with no natal kick could not be formally excluded, an asymmetric supernova explosion was found to be most likely, imparting a kick of 45--115\,km\,s$^{-1}$ to the binary, and giving rise to an eccentric orbit in the plane of the Galaxy.

In the case of XTE J1118+480, an even more compelling case for a natal kick could be made from the measured proper motion \citep{Mir01}.  The derived space velocity of 145\,km\,s$^{-1}$ relative to the LSR implied that the system was on a halo orbit, consistent with either an extraordinarily large natal kick, or formation in a globular cluster \citep[although the latter explanation was subsequently ruled out by the supersolar chemical abundances of the donor star;][]{Gon06}.  By supplementing this information with the known system parameters (component masses, orbital period, donor star properties), \citet{Fra09} were able to demonstrate that this system must have been formed with a natal kick of between 80 and 310\,km\,s$^{-1}$ \citep[see also][]{Gua05}.

In contrast to these low-mass X-ray binaries, the high-mass system Cygnus X-1 was measured to have a relatively low proper motion \citep{Che98,Mir03,Rei11}.  While the high mass of the companion should reduce the recoil velocity of the system, the observed proper motion can be explained without an asymmetric natal kick, either by symmetric mass ejection in a supernova \citep{Nel99}, or via direct collapse into a black hole \citep{Mir03}.  More detailed modelling was performed by \citet{Won12}, who used the more recent observational constraints of \citet{Rei11} and \citet{Oro11}, and were able to place an upper limit of 77\,km\,s$^{-1}$ on the natal kick velocity.

\begin{figure*}
\begin{center}
\includegraphics[width=\textwidth]{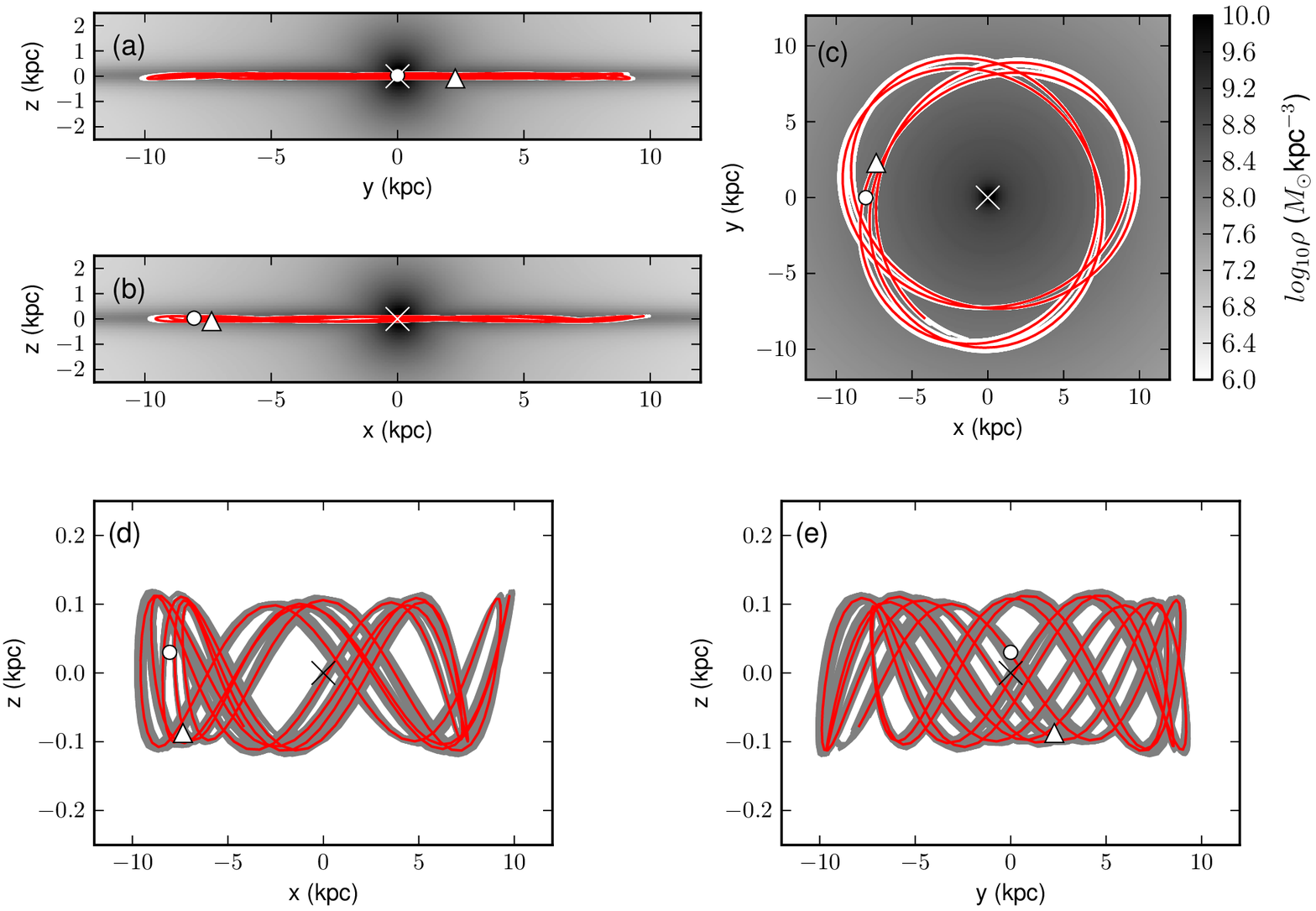}
\caption{Tracing back the trajectory of V404 Cygni through the potential of the Galaxy over the past 1\,Gyr (adapted from fig.~3 of \citealt{Mil09a}, using the updated astrometric parameters of \citealt{Mil09}).  The white circle marks the current position of the Sun, and the white triangle the current position of V404 Cygni.  The red trace shows the past trajectory for the best-fitting astrometric parameters, with the white (a,b,c) or grey (d,e) traces showing the range of possible trajectories within the $1\sigma$ uncertainties.  In panels a--c, the greyscale shows the mass density \citep[assuming the Galactic potential of][]{Joh95}.  Panels (c) and (d) show zoomed-in versions of (b) and (a), respectively.  In its orbit around the Galactic centre (marked with a cross), the vertical trajectory of V404 Cygni never reaches more than $\sim 110$\,pc above the Galactic Plane.}\label{fig:potential}
\end{center}
\end{figure*}

The only other measured black hole proper motions also suggest relatively small natal kicks.  \citet{Dha07} determined the proper motion of GRS 1915+105, which, combined with the best available systemic radial velocity of $-3\pm10$\,km\,s$^{-1}$ \citep{Gre01}, they used to determine its peculiar velocity as a function of the unknown source distance.  The proper motions of the jet ejecta during outbursts imply a maximum source distance of 11--12\,kpc \citep{Mir94,Fen99}, although a possible association with two IRAS sources has been used to argue for a distance of order 6\,kpc \citep{Kai04}.  \citet{Dha07} found the peculiar velocity to be minimised for a distance of 9--10\,kpc, and to be $<83$\,km\,s$^{-1}$ even for the maximum possible distance of 12\,kpc, and therefore concluded that no natal supernova kick was required.  Similarly, V404 Cygni \citep[$M=9^{+0.2}_{-0.6}M_{\odot}$][]{Kha10}, was found to have a peculiar velocity of 65\,km\,s$^{-1}$ \citep{Mil09,Mil09a}, which could be explained purely by a recoil kick from a natal supernova.

The measured astrometric parameters of these five black hole systems are presented in Table~\ref{tab:pms}.  However, a comparison of their peculiar velocities and the significance of any correlation with black hole mass is complicated by the differing assumptions made regarding the distance of the Sun from the Galactic Centre ($R_0$), the rotational velocity of the Galaxy ($\Theta_0$), and the solar motion with respect to the LSR, ($U_{\odot}$,$V_{\odot}$,$W_{\odot}$).  Using the values of $R_0=8.05\pm0.45$\,kpc and $\Theta_0=238\pm14$\,km\,s$^{-1}$ determined by \citet{Hon12}, and the solar motion of ($U_{\odot}$,$V_{\odot}$,$W_{\odot}$) = ($11.1^{+0.69}_{-0.75}$,$12.24^{+0.47}_{-0.47}$,$7.25^{+0.37}_{-0.36}$)\,km\,s$^{-1}$ measured by \citet{Sch10}, we have therefore applied the transformations of \citet{Joh87} to determine the full three-dimensional space velocity of each system, and used this to derive their peculiar velocities (Table~\ref{tab:vpec}).

\begin{figure}
\begin{center}
\includegraphics[width=\columnwidth]{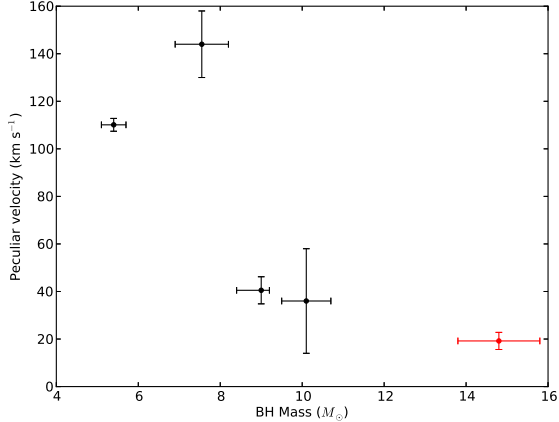}
\caption{Inferred peculiar velocity as a function of black hole mass.  Black points denote low-mass X-ray binaries, and the red point represents the high-mass X-ray binary Cygnus X-1.  A larger sample is required to make robust inferences about any potential correlation between black hole (or companion) mass and natal kicks.}\label{fig:kicks}
\end{center}
\end{figure}

With such a small sample, it is not possible to conclusively determine whether the kick velocity correlates with compact object (or total system) mass (i.e.\ to discriminate between momentum-conserving and velocity-conserving kicks), or whether there is a clear mass dichotomy between black holes forming by direct collapse and those forming in a natal supernova.  The masses of the black holes with accurate astrometric data are given in Table~\ref{tab:Mbh}, and plotted against derived peculiar velocity in Figure~\ref{fig:kicks}.  While the three lowest peculiar velocities are associated with the three highest-mass black holes, we note that more recent Bayesian methods of constraining the black hole masses \citep{Far11,Kre12} suggest that XTE J1118+480 and V404 Cyg have almost identical black hole masses.  The difference in their peculiar velocities suggests that kick velocity and black hole mass may not be directly related. 

\begin{table}
\caption{Measured black hole masses and peculiar velocities.} 
\begin{center}
\begin{tabular}{@{}cccc@{}}
\hline \hline
Source & $M_{\rm BH}$ & $v_{\rm pec}$ & References\\
& ($M_{\odot}$) & (km\,s$^{-1}$) &\\
\hline
XTE J1118+480 & 6.9--8.2 & $144\pm14$ & [1]\\
GRO J1655-40 & $5.4\pm0.3$ & $110\pm3$ & [2]\\
GRS 1915+105 & $10.1\pm0.6$ & $36\pm22$ & [3]\\
Cyg X-1 & $14.8\pm1.0$ & $19\pm4$ & [4]\\
V404 Cyg & $9.0^{+0.2}_{-0.6}$ & $40\pm6$ & [5]\\
\hline \hline
\end{tabular}\label{tab:Mbh}
\end{center}
\tabnote{Peculiar velocities taken from Table~\ref{tab:vpec}.  References for the black hole mass: [1] \citet{Kha13}; [2] \citet{Bee02}; [3] \citet{Stee13}; [4] \citet{Oro11}; [5] \citet{Kha10}.}
\end{table}

\subsubsection{Neutron stars}

Although neutron stars in X-ray binaries should have formed in a supernova explosion, their existence in a binary system implies that their natal kicks should typically have been lower than those deduced for the radio pulsar population, since a sufficiently strong natal kick would unbind the binary.  Indeed, the population of recycled pulsars (spun up due to accretion from a binary companion) is observed to have a significantly lower mean space velocity than that of normal pulsars \citep{Hob05}.

Some of the first X-ray binary proper motions were determined from Hipparcos data for a sample of high-mass neutron star systems \citep{Che98}.  They showed that the mean transverse velocities of Be/X-ray binaries were lower than those of supergiant systems.  Optical observations have also been used to determine the proper motion of Cen X-4 \citep[][see Table~\ref{tab:pms}]{Gon05}.

Owing to their intrinsically fainter radio emission \citep{Mig06}, only a handful of neutron star X-ray binaries are accessible to the higher precision astrometry possible using VLBI.  Of the five systems with VLBI proper motions, two (Sco X-1 and Cyg X-2) are Z-sources, persistently accreting at or close to the Eddington rate, and showing resolved radio jets \citep{Fom01,Spe13}.  The atoll source, Aql X-1, was observed during two of its transient outbursts \citep{Mil10,Tud13}, when it showed only marginal evidence for resolved jets.  The remaining two systems are gamma-ray binaries (LS 5039 and LSI +61$^{\circ}$303), whose radio emission is instead likely to arise from a collision between the relativistic wind of a pulsar and the stellar wind of its companion star.

Of these systems, Sco X-1, Cen X-4 and LS 5039 were all deemed to have undergone a strong natal kick at formation \citep{Mir03a,Gon05,Mol12a}, and despite its low peculiar velocity, the high eccentricity of LSI +61$^{\circ}$303 is a strong indication that it should also have received an asymmetric kick in the natal supernova \citep{Dha06}.  This leaves Aql X-1 as the only neutron star system in our sample (Table~\ref{tab:vpec}) that is unlikely to have received a strong natal kick.

\subsubsection{Neutron stars and black holes: a comparison}

Although the Galactic distribution of neutron star and black hole X-ray binaries suggests similar natal kicks for neutron star and black hole systems \citep{Jon04}, we can directly test this by determining whether the neutron star and black hole peculiar velocities in Table~\ref{tab:vpec} are drawn from the same underlying distribution.  Despite the three systems with the highest peculiar velocities being neutron stars, a Kolmogorov-Smirnov test suggests that the null hypothesis cannot be ruled out at better than the 63\% level.  Thus there is no statistically significant difference between the current samples.  However, those samples are small, and were selected primarily (but not exclusively) on the basis of radio brightness.  Thus, a meaningful test of this hypothesis requires further astrometric measurements.

\subsection{Birthplaces}

For the youngest systems (i.e.\ the high-mass X-ray binaries), detailed position and velocity information can allow us to determine the birthplace of the compact object.  Based on the low peculiar velocity of Cygnus X-1 and the similarities between its proper motion and that of the nearby star cluster Cygnus OB3, \citet{Mir03} suggested that Cygnus X-1 had originated in this star cluster.  However, in the case of Cygnus X-3, an unknown compact object in orbit with a WN7 Wolf-Rayet star \citep{vanKer92}, the high mass loss rate in the stellar wind has to date precluded the identification of optical lines from the disc or companion star, such that the systemic radial velocity is poorly constrained ($|\gamma|<200$\,km\,s$^{-1}$). Nevertheless, \citet{Mil09b} used archival VLA and VLBA data to determine the proper motion of the system, and inferred a peculiar velocity in the range 9--250\,km\,s$^{-1}$.  Although the Wolf-Rayet companion implies that the system must be relatively young, no potential progenitor star cluster could be identified owing to the uncertain systemic radial velocity and the high extinction along the line of sight.

Astrometric measurements have also shed light on the origin of the persistent, super-Eddington system SS 433.  It is offset a few pc to the west of the centre of the W50 nebula \citep{Loc07} that is believed to be the supernova remnant arising from the creation of the compact object.  Its measured three-dimensional space velocity is of order 35\,km\,s$^{-1}$, oriented back towards the Galactic Plane, suggesting that the original binary system was originally ejected from the Galactic Plane.  The compact object progenitor then underwent a supernova explosion within the past $10^5$\,yr, giving rise to a small natal kick that can account for the current peculiar velocity and offset from the centre of W50 \citep{Loc07}.  

The proximity of the neutron star system Circinus X-1 to the supernova remnant G321.9-0.3 led to similar suggestions, that this SNR was the birthplace of the X-ray binary \citep{Cla75}.  However, the HST upper limit on the proper motion of the system subsequently ruled out this scenario \citep{Mig02}. This conclusion was recently confirmed by the detection of a faint X-ray nebula surrounding the X-ray binary, identified (together with the associated radio nebula) as the supernova remnant from the formation of the neutron star, placing an upper limit on its age of 4600 years \citep{Hei13}.

Assuming that black holes formed in the Galactic Plane \citep[as inferred in several cases from the chemical abundances of their secondary stars; e.g.][]{Gon08}, accurate three-dimensional space velocities for black holes can also provide lower limits on their ages. By tracing their trajectories back in time in the Galactic potential (e.g.\ Fig.~\ref{fig:potential}), the times at which they crossed the Plane can be determined.  The most recent crossing that also satisfies constraints from binary evolution modelling then provides a lower limit on the age of the black hole \citep[e.g.][]{Fra09}.

Finally, for those X-ray binaries detected within globular clusters, the natal kicks must have been sufficiently small for the systems to remain bound to the host cluster \citep[see, e.g.][]{Pfa02a}.  Astrometric proper motion measurements could both confirm an association with the cluster, and allow us to probe the movements of the systems within the cluster potential, improving our understanding of the intracluster dynamics.

\section{Astrometric residuals}
\label{sec:residuals}

Having fit a set of positional measurements to determine the proper motion and parallax of the target source, the astrometric residuals contain additional information on the size scales of both the jets and the binary orbit, which can be probed using sufficiently precise measurements.

\begin{figure*}[!t]
\begin{center}
\mbox{\subfigure[VLBA image of the compact jet in Cyg X-1, from a hard state observed on 2009 July 14th.  The observing frequency was 8.4\,GHz.]{\epsfig{figure=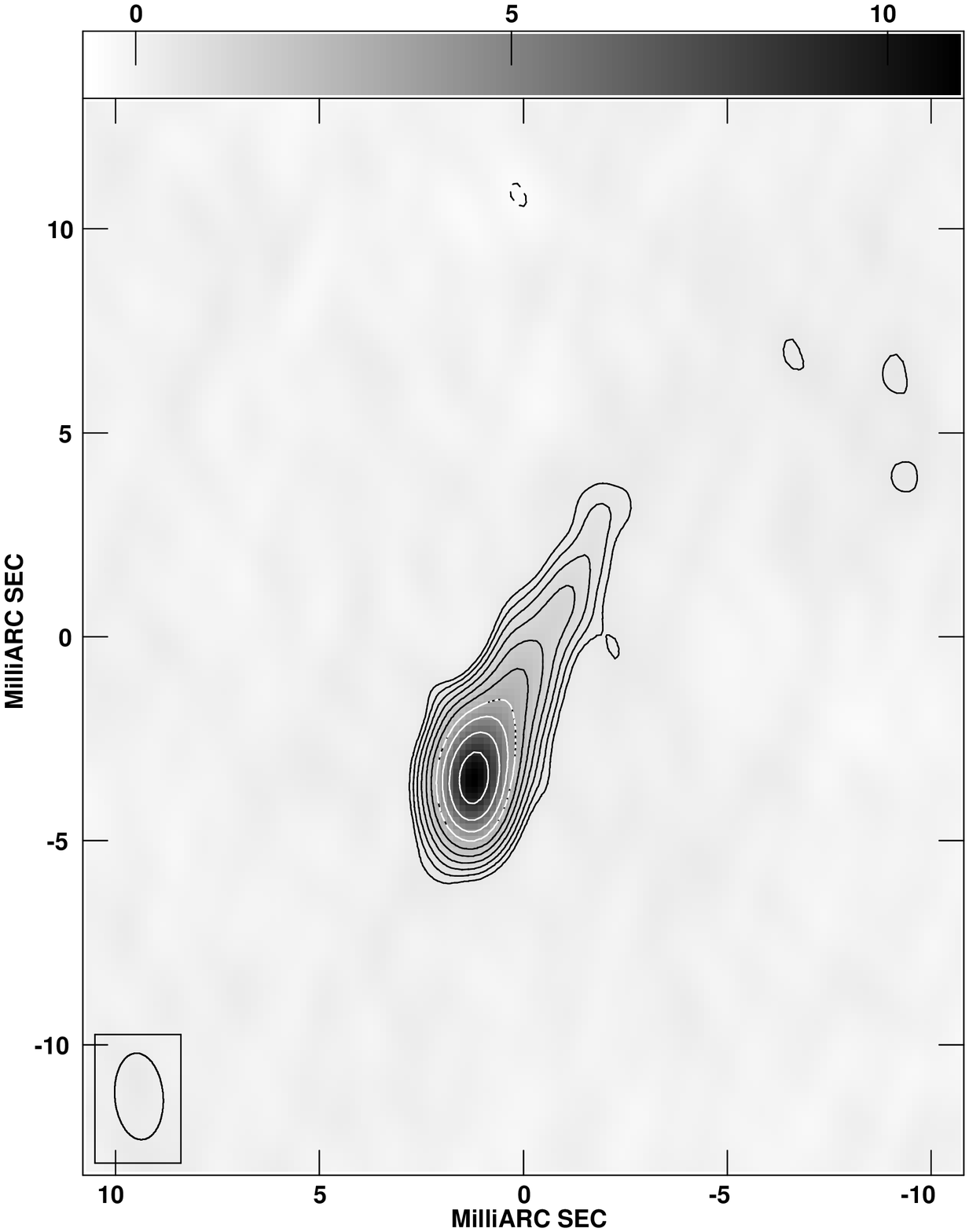,height=0.31\textwidth}}\quad
  \subfigure[Orbital phase dependence of the position of Cyg X-1.  Having subtracted off the proper motion and parallax signatures \citep{Rei11}, this plot shows the astrometric residuals perpendicular to the jet axis during a hard-to-soft state transition in 2010 July.  Numbering shows the different epochs of observation.  Modelled orbital signatures are shown for counterclockwise (dotted) and clockwise (dashed) orbits, using the size predicted from \citet{Oro11}.]{\epsfig{figure=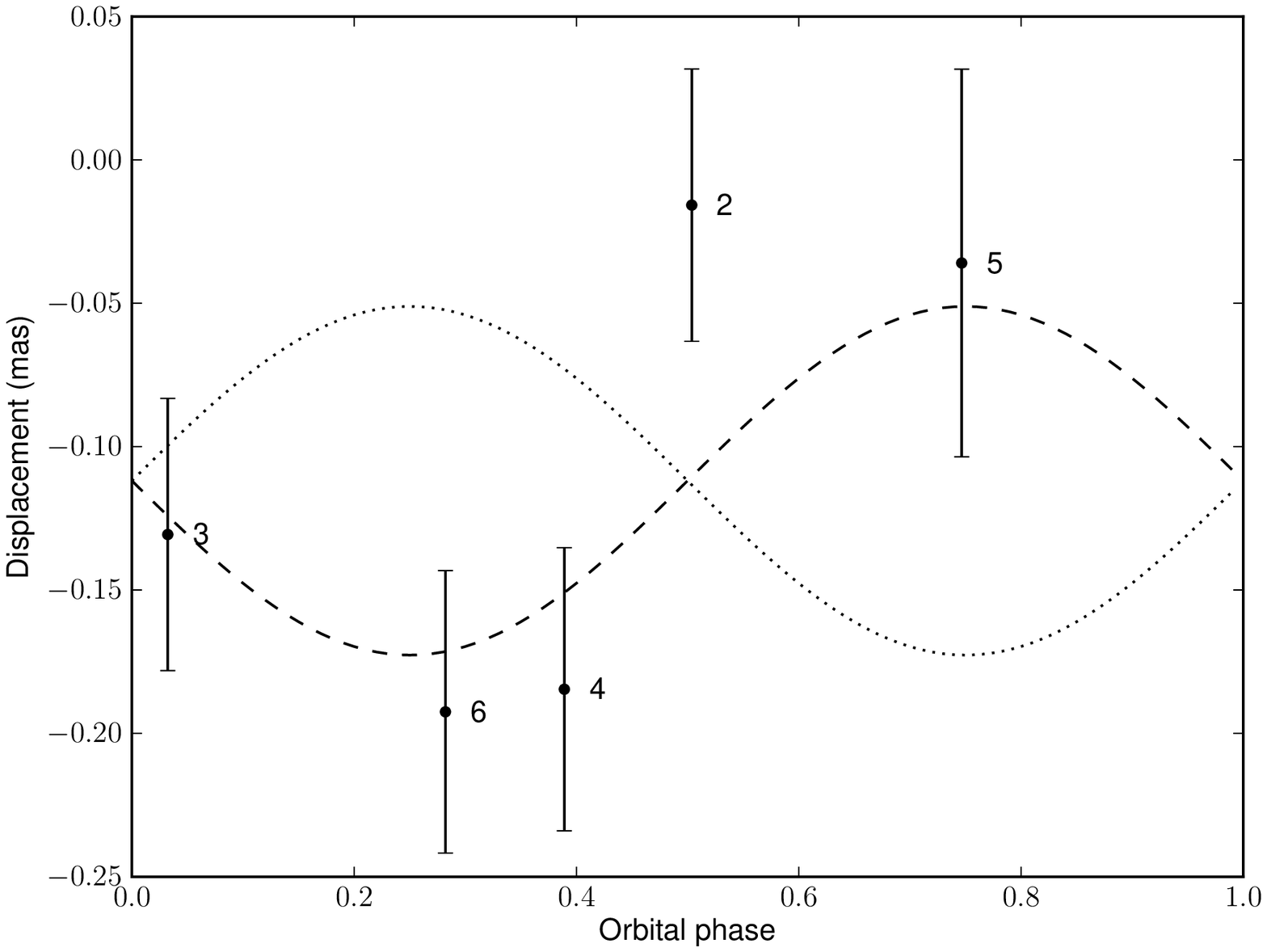,height=0.28\textwidth,bbllx=5,bblly=5,bburx=546,bbury=392,clip=}}\quad
  \subfigure[Alignment of the astrometric residuals from (b) with the known jet axis, after subtracting the known proper motion, parallax and orbital motion signatures \citep{Rei11}.  Dashed line shows the position angle of the resolved VLBI jet in (a).  The offset between frequencies is caused by frequency-dependent structure in the phase reference calibrator (adapted with permission from fig.~2 in \citealt{Rus12}).]{\epsfig{figure=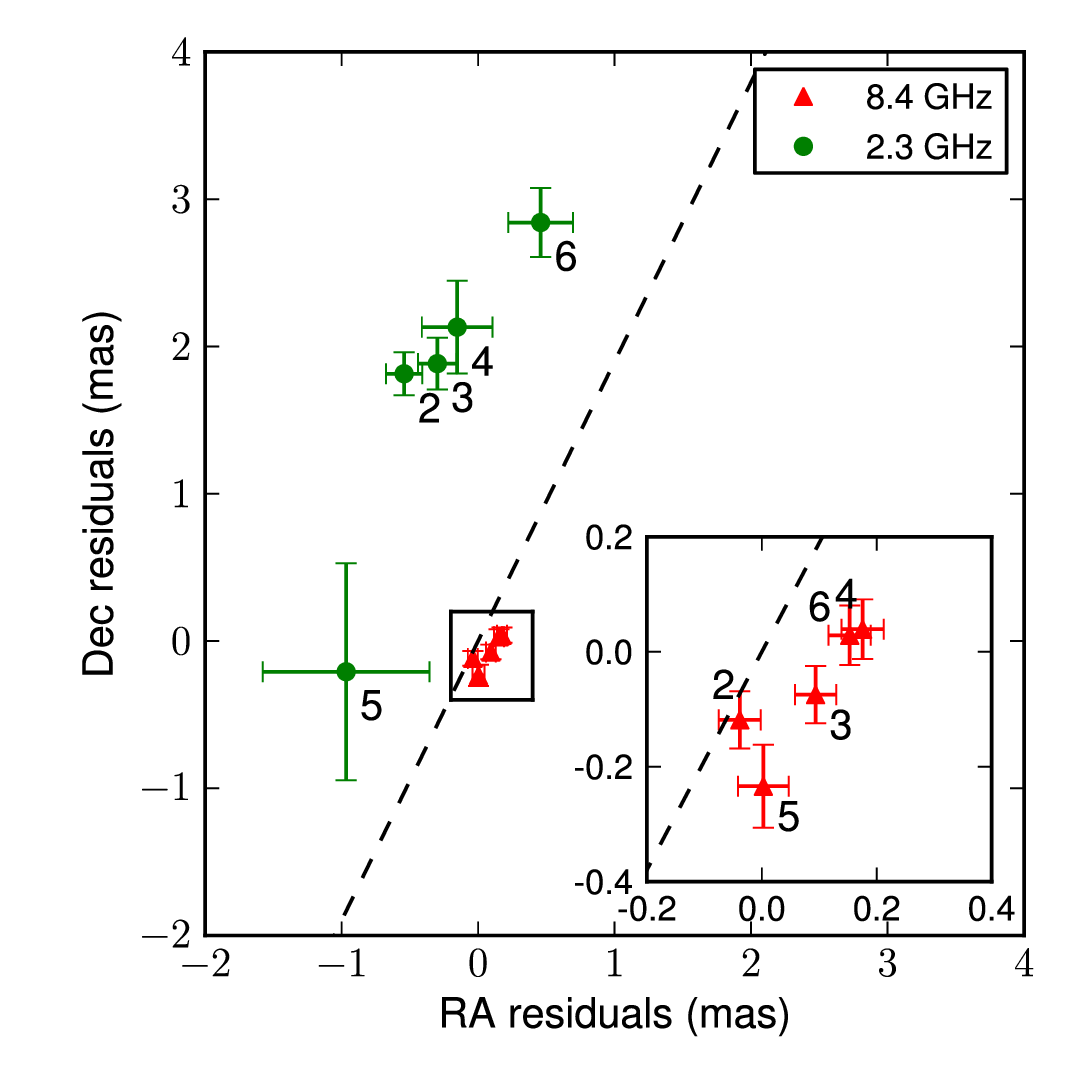,height=0.31\textwidth}}}
\caption{Astrometric residuals in Cygnus X-1.}
\label{fig:residuals}
\end{center}
\end{figure*}

\subsection{Orbital phase-resolved astrometry}

The radio emission detected by VLBI typically arises from a steady, compact, partially self-absorbed radio jet \citep{Bla79}, likely launched from a few tens of gravitational radii from the black hole.  With sufficiently high-precision astrometry, the orbital signature of the black hole around its companion can be determined.  The size of this orbital signature is given by
\begin{equation}
r_{\rm BH} = \frac{M_{\rm d}}{(M_{\rm BH}+M_{\rm d})^{2/3}}\left(\frac{GP^2}{4\pi^2}\right)^{1/3},
\end{equation}
where $P$ is the orbital period, and $M_{\rm BH}$ and $M_{\rm d}$ are the masses of the black hole and donor star, respectively.  This implies that such measurements are likely to be successful only for systems with high-mass companions and long orbital periods.  Should this be feasible, however, it can provide independent constraints on the system parameters.  In the only example to date, \citet{Rei11} were able to measure the orbital signature of the black hole in Cygnus X-1, with the reduced $\chi^2$ values favouring a clockwise orbit (see also Figure~\ref{fig:residuals}b).  In this case, the system parameters of \citet{Oro11} were supplied to determine the magnitude of the orbital signature, but with sufficiently high-precision measurements, astrometric data could be used to constrain this independently.

In neutron star systems, where the mass ratio $q_1 = M_{\rm x}/M_{\rm d}$ is smaller, the orbital signature should be easier to determine.  In a sequence of 12 VLBA observations sampling the full 26-day orbit of the gamma-ray binary LSI +61$^{\circ}$303, \citet{Dha06} showed that the measured source position traced out an elliptical locus on the plane of the sky, interpreted as the orbital signature of the source.  However, the measured size of the ellipse was found to be significantly larger at 2.3\,GHz than at 8.4\,GHz, and in both cases was much larger than the size of the orbit inferred from the measured system parameters.  Furthermore, the position angle of the 2.3-GHz emission trailed that of the higher-frequency emission, leading \citet{Dha06} to suggest that they were observing emission from an extended cometary tail that trailed the orbit of the neutron star, hence favouring a pulsar wind origin for the observed radio emission (as in PSR B1259-63) over a precessing radio jet.  Although a reanalysis of the same data by \citet{Mas12} was used to argue for the precessing jet scenario, follow-up observations by \citet{Mol12} have provided further evidence for the pulsar wind model.

\subsection{Core shifts and the size scale of the jets}
\label{sec:core_shifts}

Having subtracted off the parallax, proper motion and orbital motion signatures from the measured source positions, the astrometric residuals should lie along the axis of the radio jets (Figure~\ref{fig:residuals}c).  In a classical partially self-absorbed jet, we see emission from the surface where the optical depth is unity at the observing frequency.  This implies that lower-frequency emission arises from further downstream, giving rise to a core shift, as frequently observed in AGN \citep[e.g.][]{Lob98,Kov08}.  Measurement of the core shift as a function of frequency is a well-established technique in AGN, where the lack of a moving source makes such measurements easier, and can be used to determine the structure of the jet.

Since astrometric positions are measured relative to the assumed position of a nearby background calibrator, which may differ at different frequencies (owing to its own core shift), core shift measurements require careful astrometry.  For the gamma-ray binary LSI +61$^{\circ}$303, \citet{Mol12} has shown that when using multiple extragalactic calibrator sources it is possible to disentangle the core shifts of both the calibrators and the target source, although such a technique has not to date been applied to any other X-ray binaries.

\subsection{Jet orientation and extent}

The compact jets seen in the hard and quiescent states, which form the astrometric targets for VLBI observations, are known to be variable \citep[e.g.][]{Mil09}.  Small-scale flaring events can arise from changes in the velocity, magnetic field strength, or electron density at the base of the jets.  This can affect the position of the $\tau=1$ surface, causing it to move up or downstream along the jet axis.  The astrometric residuals from such compact jets at a range of different brightnesses can therefore be used to determine the orientation of the jet axis and the extent of the jets as a function of frequency, once the proper motion, parallax and orbital signatures have been removed.  \citet{Rus12} used the previously-determined astrometric parameters of Cygnus X-1 \citep{Rei11} to infer the existence of a remnant compact jet from the astrometric residuals as the source began a transition to its softer X-ray spectral state (when the compact jets are usually believed to be quenched).  The astrometric residuals were seen to align with the well-known jet axis from VLBI imaging \citep[][see also Fig.~\ref{fig:residuals}a]{Sti01}, and showed more scatter at 2.3\,GHz than at 8.4\,GHz, easily explained in the scenario where the jets are more extended at lower frequencies owing to the $\tau=1$ surface being further downstream (Figure~\ref{fig:residuals}c).

\section{Future prospects}
\label{sec:future}

The majority of persistent, bright X-ray binaries have already been the targets of astrometric observations (Table~\ref{tab:pms}).  Since astrometric observations with VLBI are only possible in the hard spectral state seen at the beginning and end of an outburst, and for the closest or brightest quiescent systems \citep{Gal06,Mil11}, relatively few easily-accessible targets remain.  While proper motions can be measured from hard state observations at the start and end of a single X-ray binary outburst \citep[e.g.][]{Mir01}, accurate parallax measurements require observations across the full parallax ellipse, particularly at the times of maximum and minimum parallax displacement.  Since the timing of X-ray binary outbursts is unpredictable, accurate parallax measurements will therefore be difficult, except for recurrent systems with a high duty cycle, such as Aql X-1, H1743-322 or GX339-4.  Indeed, the feasibility of determining a parallax from the recurrent outbursts of a transient source has recently been demonstrated for the dwarf nova SS Cygni \citep{Mil13}.  The only alternative would be for array sensitivity enhancements to make fainter quiescent systems accessible to VLBI.

Such sensitivity improvements to existing VLBI arrays are either being planned or are already underway, and will both improve the astrometric accuracy of existing measurements and extend the range of possible targets to fainter systems.  The recent trend to increasing recording rates (rates up to 2048\,Mbps are now standard) not only improves the sensitivity of a VLBI array, but also permits the use of fainter, closer phase referencing calibrators.  This improves the success of the phase referencing process and reduces the astrometric systematics, which scale linearly with calibrator-target separation \citep{Pra06}.  With the option to simultaneously correlate on multiple phase centres at once via $uv$-shifting, software correlators \citep[e.g.][]{Del07,Del11} have made it possible to find in-beam calibrators for the majority of low-frequency ($\lesssim 1.4$\,GHz) VLBI observations.  As well as improving the accuracy of the phase transfer due to the proximity of target and calibrator source, this reduces the slewing and calibration overheads associated with the observation, allowing more time to be spent on the science target.

While these improvements should increase the sensitivity of VLBI arrays by factors of a few, only the large increase in collecting area provided by connecting the SKA to existing VLBI arrays will permit the extension of astrometric studies to a significant number of faint, quiescent systems.  Also, by enabling the detection of faint radio emission from extragalactic black holes (either X-ray binaries or ULXs), it could, given a sufficiently long time baseline, allow the measurement of the proper motions of the most luminous black holes in nearby galaxies \citep[or even their ejecta, as tentatively reported for an exceptionally bright transient in M82;][]{Mux10}.  Although the details are still to be determined, a VLBI capability is envisaged in the SKA baseline design, and the main science drivers have been presented by \citet{God12}.  The high sensitivity of the SKA could also allow the detection of radio emission from isolated black holes accreting via Bondi-Hoyle accretion from the interstellar medium \citep{Mac05}, and \citet{Fen13} suggested that astrometric observations could identify such systems via their high proper motions of a few tens to hundreds of mas\,year$^{-1}$, corresponding to velocities of up to several tens of kilometres per second for black holes at a distance of 100\,pc.

Increasing the frequency of VLBI observations into the sub-millimetre band improves the resolution and hence the astrometric accuracy of the observations.  The advent of the Event Horizon Telescope \citep{Doe09}, combining existing and planned sub-millimetre telescopes \citep[including the phasing up of the Atacama Large Millimeter Array; see][]{Fis13} will enable sensitive VLBI observations at millimetre and sub-millimetre wavelengths.  With an astrometric precision of a few microarcseconds, this would allow us to resolve the orbits of binary systems, and potentially even detect the thermal emission from the donor star.  By tracking the orbits of both components, the system parameters could be constrained with unprecedented accuracy.

Moving from the radio to the optical band, the GAIA astrometric mission \citep{Per01} promises to revolutionise Galactic astrometry.  With the aim of measuring astrometric parameters for $10^9$ stars, complete to $V=20$, this mission will measure geometric parallaxes to an accuracy of 11\,$\mu$as at $G=15$, degrading to 160\,$\mu$as at $G=20$.  For stars brighter than $G=$17--18, its astrometric accuracy will thus rival or exceed that currently achievable with typical VLBI observations.  Although most transient X-ray binaries spend the majority of their duty cycles in quiescence with $V>20$ \citep[e.g.][]{Sha99b}, any high-mass or persistent systems (such as Cygnus X-1), as well as the brightest quiescent systems (e.g.\ V4641 Sgr, 4U1543-47, GRO J1655-40) should be accessible, and we estimate at least $\sim14$ potential targets amongst the known black hole and black hole candidate systems (D.~M. Russell, priv.\ comm.).  However, for highly-extincted systems in the Galactic Plane, VLBI radio observations will remain the astrometric technique of choice.

\section{Summary}

Astrometric observations are among the most fundamental astronomical measurements.  Over the past two decades, high-precision astrometric observations with VLBI arrays have made it possible to determine the proper motions of radio-emitting X-ray binary systems across the Galaxy, and, in a few cases, to determine model-independent distances via geometric parallax.  These measurements can probe a range of fundamental physics, from black hole formation mechanisms to the neutron star equation of state, the existence of event horizons, and the spin-powering of black hole jets.  Important constraints on system parameters can also be derived from accurate source distances, and even from astrometric residuals, after subtracting the parallax and proper motion signatures from the measured positions.

Such astrometric observations are restricted to systems with detectable, compact radio emission that is causally connected to the central binary system, i.e.\ X-ray binaries in their hard or quiescent states.  For all but the brightest quiescent systems, this requires triggered VLBI observations in the hard X-ray spectral states seen during the rise and decay of X-ray binary outbursts.  The proper motion can be measured over the course of a typical few-month transient event, and for recurrent transients, observations over several outbursts can determine the parallax.  While high-impact results can be derived from astrometry of individual systems, a large sample of proper motions is required to place useful observational constraints on black hole formation.  With only 1--2 black hole X-ray binary outbursts per year \citep{Dun10}, it is therefore important to take advantage of every opportunity to make VLBI observations of X-ray binaries in their hard states, particularly given the sessional nature of both the European VLBI Network and the Australian Long Baseline Array.  Although GAIA will significantly extend the sample of X-ray binaries with measured astrometric parameters, the optical faintness of quiescent X-ray binaries and the extinction in the Galactic Plane implies that VLBI will continue to play an important role in such astrometric studies.

\begin{acknowledgements}
JCAMJ acknowledges support from Australian Research Council Discovery Grant DP120102393, and thanks Tom Maccarone, Peter Jonker and the anonymous referee for insightful comments on the manuscript, Dave Russell for useful discussions, and Andrzej Zdziarski for pointing out an inconsistency in the original version of Fig.~\ref{fig:potential}.  This work has made use of NASA's Astrophysics Data System.
\end{acknowledgements}

\bibliographystyle{apj}
\bibliography{millerjones_astrometry}



\end{document}